\documentclass[referee]{raa}            

\usepackage{graphicx,times}             
\usepackage{multirow}
\usepackage{booktabs} 
\usepackage{natbib}
\usepackage{amssymb,amsmath}
\usepackage{algorithm,algorithmic}
\usepackage{rotating}
\usepackage{lscape}
\bibpunct{(}{)}{;}{a}{}{,}

\usepackage{threeparttable}
\usepackage[pagebackref=true]{hyperref}

\begin{document}

  \title{An in-depth exploration of LAMOST Unknown spectra based on density clustering}

   \volnopage{Vol.0 (20xx) No.0, 000--000}      
   \setcounter{page}{1}          

   \author{Hai-Feng Yang 
      \inst{1,4}
   \and Xiao-Na Yin
      \inst{1,4}
   \and Jiang-Hui Cai
      \inst{1,2,4}
   \and Yu-Qing Yang
      \inst{4}
   \and A-Li Luo
      \inst{3}
   \and Zhong-Rui Bai
      \inst{3}
      \and Li-Chan Zhou
      \inst{4}
   \and Xu-Jun Zhao
      \inst{4}
   \and Ya-Ling Xun
      \inst{4}
   }

   \institute{Shanxi Key Laboratory of Big Data Analysis and Parallel Computing, Taiyuan University of Science and Technology, Taiyuan 030024, China; \\
        \and
             School of Computer Science and Technology, North University of China, Taiyuan 030051, China; {\it jianghui@tyust.edu.cn}\\
        \and
             National Astronomical Observatories, Chinese Academy of Sciences, Beijing 100101, China\\
        \and
             School of Computer Science and Technology, Taiyuan University of Science and Technology, Taiyuan 030024, China; \\
\vs\no
   {\small Received 20xx month day; accepted 20xx month day}}

\abstract{LAMOST (Large Sky Area Multi-Object Fiber Spectroscopic Telescope) has completed the observation of nearly 20 million celestial objects, including a class of spectra labeled `Unknown'. Besides low signal-to-noise ratio, these spectra often show some anomalous features that do not work well with current templates. In this paper, a total of 638,000 `Unknown' spectra from LAMOST DR5 are selected, and an unsupervised-based analytical framework of `Unknown' spectra named SA-Frame (Spectra Analysis-Frame) is provided to explore their origins from different perspectives. The SA-Frame is composed of three parts: NAPC-Spec clustering, characterization and origin analysis. First, NAPC-Spec(Nonparametric density clustering algorithm for spectra) characterizes different features in the "unknown" spectrum by adjusting the influence space and divergence distance to minimize the effects of noise and high dimensionality, resulting in 13 types. Second, characteristic extraction and representation of clustering results are carried out based on spectral lines and continuum, where these 13 types are characterized as regular spectra with low S/Ns, splicing problems, suspected galactic emission signals, contamination from city light and un-gregarious type respectively. Third, a preliminary analysis of their origins is made from the characteristics of the observational targets, contamination from the sky, and the working status of the instruments. These results would be valuable for improving the overall data quality of large-scale spectral surveys.
}

   \authorrunning{H.-F. Yang, X.-N. Yin, J.-H. Cai et al.}            
   \titlerunning{An in-depth exploration of Unknown spectra}  

   \maketitle

%
%
\section{Introduction}           
\label{sect:intro}

The Large Sky Area Multi-Object Fiber Spectroscopic Telescope (LAMOST) is a special quasi-meridian reflecting Schmidt telescope located in Xinglong Station of National Astronomical Observatory, China \citep{cui2012large}. Totally 9,026,365 targets are released in LAMOST DR5, including 8,183,160 stars, 152,863 galaxies, 52,453 quasars, and 637,889 `Unknown' objects. These `Unknown' objects are the spectra that can not be classified by LAMOST 1-D pipeline. However, they are valuable for studying their origins  and discovering unknown objects. Therefore, the main motivations of this work are listed as follows.
 
$\bullet$ The exploration of `Unknown' spectra is beneficial to improve the spectral processing techniques. Meanwhile, it can distinguish between avoidable and unavoidable factors and improve data product quality.

$\bullet$ Limited by the completeness of the template, some rare spectra may be hidden in the `Unknown' dataset. Studying these samples may help to discover and understand the rare features. 

In recent years, many studies have been carried out to obtain valuable information from LAMOST `Unknown' spectra, and some special and rare targets are mined and identified from LAMOST spectra. For example, from `Unknown' spectra, \citet{huo2017quasars} identified high-redshift quasars  by eyeball inspection, \cite{guo2019recognition} recognized M-type stars, \cite{li2018carbon} searched for carbon stars using machine learning algorithms. \cite{10.1093/mnras/sty805} present the DR5 catalogue of white dwarf–main sequence binaries from the LAMOST, part of which also originate from the `Unknown' spectra. Meanwhile, some effective methods have been proposed in the study of `Unknown' spectra. For example, neural network-based feature extraction methods \citep{wang2017new,bu2015restricted} are used in defective spectra recovery. In addition, in order to re-identify `Unknown' spectra, \cite{zheng2020classification} used a deep learning-based classification method to classify `Unknown' spectra in LAMOAT DR6 as candidates for galaxies, QSOs and stars, then classified the candidate stellar spectra into subclasses from O to M. From the above analysis, most of the studies focused on discovering valuable information from the spectra or effectively avoiding interference on the research goals.

 It is also meaningful to explore the `Unknown' spectra to improve the data quality and promote spectral processing techniques by using new data mining methods \citep{10.1093/mnras/stac2975,10.1093/mnras/stac3292}. This paper designs an unsupervised-based analytical framework and performs a detailed analysis of LAMOST `Unknown' spectra. The main contributions are as follows:

$\bullet$ A framework named SA-Frame is built for the analysis of `Unknown' spectra. NAPC-Spec is designed for the characterization of diverse features in `Unknown' spectra. NAPC-Spec deploys influence space and divergence distance to minimize the influence of noise and high dimensionality.

$\bullet$ The `Unknown' spectra from LAMOST DR5 are explored by SA-Frame. These spectra are classified into 13 types and characterized as regular spectra with low S/Ns, splicing problems, suspected galactic emission signals, contamination from city light and un-gregarious type. In addition, the source of these `Unknown' spectra is analyzed from the characteristics of the observational targets, contamination from the sky, and the working status of the instruments.

The article is organized as follows. In Section 2, we describe the data selection. In Section 3, we give the `Unknown' spectra analysis framework. Section 4 presents the clustering results and analysis for `Unknown' spectra. Origin analysis of `Unknown' spectra is provided in Section 5. Finally, the discussion is carried out in Section 6. 

\section{Data selection}
The spectra were marked as `Unknown' because the spectra do not match well with any known templates. These data may not only be low signal-to-noise ratio, but also may have quality defects, such as abnormal continuum, abnormal splicing, unclear spectral line characteristics and missing data in some bands. LAMOST has completed 9 data releases since its official survey in October 2011 \citep{10.1117/12.2055695}. In this paper, a total of 637,994 `Unknown' spectra were selected from LAMOST DR5 for unsupervised density-based clustering analysis \citep{CAI2023}. The mean signal-to-noise ratio of these `Unknown' spectra is 5.25. All these spectra are from the LAMOST low-resolution spectral survey, which has a resolution $R\sim 1800$ at 5500 $\AA$ and wavelength coverage of  3690$\AA$ -- 9100$\AA$ \citep{luo2022vizier}.

\section{SA-Frame: `Unknown' Spectra Analysis Framework}
\begin{figure*}
\centering
\includegraphics[width=\linewidth]{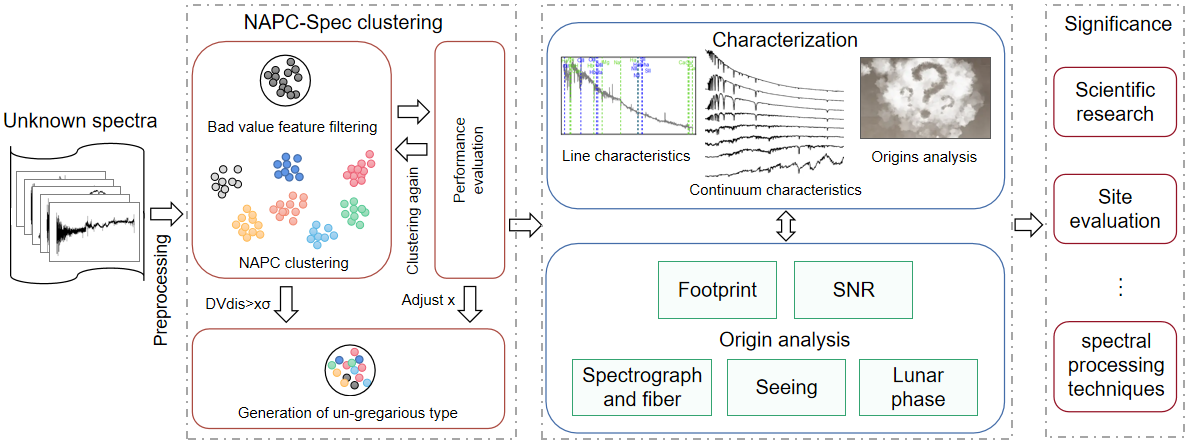}
\caption{The architecture of SA-Frame.}
\label{fig:example0} 
\end{figure*}

The main purpose of analysis on `Unknown' spectra is to trace the underlying origins through the massive data. In order to systematically explore the origin of the `Unknown' spectra, an analytical framework named SA-Frame is designed. As shown in Figure \ref{fig:example0}, SA-Frame contains three parts: NAPC-Spec clustering, characterization and origin analysis. 

\subsection{NAPC-Spec clustering algorithm}\label{sec:2-1}
An extended NAPC \citep{yang2022isbfk,yang2022density} density clustering method for spectral analysis, named NAPC-Spec, is provided for efficient clustering of `Unknown' spectra. In this algorithm, alignment, normalization and abnormal pixel elimination are performed in steps 1-3; then, put the spectra with the number of continuous bad pixels (flux=0) n\textgreater 40 into a separate type; next, different types are obtained by clustering the spectra with the NAPC algorithm; finally, the spectra with DVdis \textgreater $3 \sigma$ are filtered and placed in an un-gregarious type.
The description is detailed in Algorithm \ref{alg:1}. 

\begin{algorithm}
	\renewcommand{\algorithmicrequire}{\textbf{Input:}}
	\renewcommand{\algorithmicensure}{\textbf{Output:}}
	\caption{The NAPC-Spec algorithm}
	\label{alg:1}
	\begin{algorithmic}[1]
		\REQUIRE LAMOST `Unknown' spectral dataset $D$
		\ENSURE $k$ types
		\STATE Dimensional unification and alignment. 
		\STATE Normalization of spectral fluxes using Z-Score method.
		\STATE Abnormal pixel elimination. The anomalous pixels in the spectra are eliminated and the local spectra are obtained by employing the noise recognition and deletion algorithm from ARIS \citep{10.1145/3522592}.          
            \STATE Bad pixel values feature filtering. Put the spectra with the number of continuous bad pixels (flux=0) n\textgreater 40 into a separate type.   
		\STATE Feature extraction. The features are extracted from the abnormal-pixels-eliminated local spectra based on the influence space of ARIS. 
		\STATE Clustering the remaining spectra using the NAPC algorithm to obtain $k'$ types.
		\STATE Generation of un-gregarious type. The divergence distance DVdis from the spectra in each type to the type center is calculated, then the spectra with DVdis \textgreater $3 \sigma$ are filtered and these spectra in the same type are placed in un-gregarious type.			
		\STATE \textbf{return} $k$ ( $k=k'+2$) target types.
	\end{algorithmic} 
\end{algorithm} 

 In the Algorithm \ref{alg:1}, the divergence distance (as shown in Equation (\ref{eq:1})) is used as the similarity measurement, which can effectively avoid the similarity transfer problem of Euclidean distance in the calculation of high-dimensional data. 
\begin{equation}\label{eq:1}
{\rm DVdis}({{P}_{i}},{{P}_{j}})={\rm Eu}({{P}_{i}},{{P}_{j}})-{{e}^{-{{({\rm DV}({{P}_{i}})-{\rm DV}({{P}_{j}}))}^{2}}}}
\end{equation} 
where ${\rm DV}({{P}_{i}})=\frac{\sum\nolimits_{k=1,k\ne i}^{m}{\rm Eu({{P}_{k}},{{P}_{i}})}}{\xi *d}$ , ${P}_{k}$ , ${P}_{i}$ are any two spectra in the dataset $D$, $d$ is the diameter of the circle where these points are located, $\xi$ is the accumulation times.

In addition, the initial type center is automatically obtained by a judgment index $Z$ (Equation (\ref{eq:2})), which can improve the clustering efficiency.
\begin{equation}\label{eq:2}
{{Z}_{i}}={{\rho }_{i}}*\min ({\rm DVdis}({{P}_{i}},{{P}_{j}})),{{\rho }_{i}}<{{\rho }_{j}},0\le i,j\le n
\end{equation}
where ${\rho }_{i}$ is the local density of ${P}_{i}$, ${{\rho }_{i}}=\sum\nolimits_{j=1}^{J}{{{e}^{-(\frac{{\rm DiffDVdis}({{P}_{i}},{{P}_{j}})}{d{{c}_{i}}})}}},{{P}_{j}}\in {\rm NP}({{P}_{i}})$ , ${\rm NP}({{P}_{i}})$  is the set of neighbors for ${P}_{i}$,  $J$ is the number of elements in ${\rm NP}({{P}_{i}})$, ${\rm DiffDVdis}$ is the divergence distance difference between the two points.

\subsection{Characterization and origin analysis procedure}\label{sec:2-2}

A characterization procedure for each type based on spectral lines, continua shape, and other physical properties. First, emission/absorption line features are identified. Second, the continua are characterized according to the change trends of rising, falling and other complex variations. Third, the examination of sky-light and city lights. In particular, the composite spectra superimposed by moonlight are given special attention. In the end, other un-recognized abnormal and obvious features are described separately for follow-up study.

To explore the origins of each type of `Unknown' spectra, the footprint, S/N of the observational targets, and the observational status, including moon phase, seeing and working status of the spectrographs and fibers are then comprehensively studied.

\section{Clustering results analysis of `unknown' spectra}\label{sec:3}
This section provides an overview of the clustering results, and gives explanations regarding the type characteristics. Notably, SA-Frame is an unsupervised learning approach that is driven by the data itself, and the data with common features will theoretically be classified into a cluster.

\subsection{Overview of clustering results}
\begin{figure}[htpb]
\centering
\includegraphics[scale=0.9]{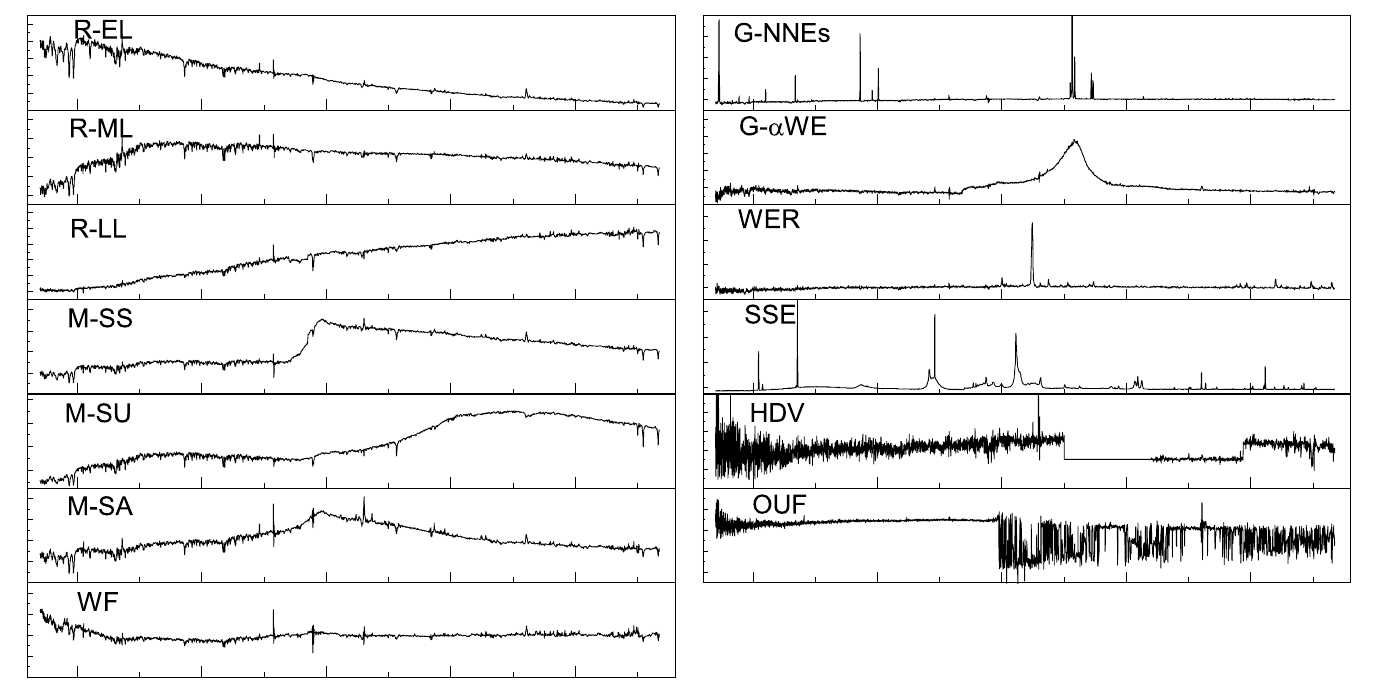}
\caption{Thumbnails of the 13 types, with the name of each type feature in the upper left corner.}
\label{fig:example1}
\end{figure}

In this paper, Algorithm \ref{alg:1} divides the `Unknown' spectra from LAMOST DR5 into 13 types. The composite spectra (the average spectra of each type) of the first 11 types and four random samples from the last 2 types are shown in Figure\ref{fig:example1}. The basic information of each type is shown in Table \ref{tab1}, including Type name, Number, Proportion, S/N (average, range), etc. Table \ref{tab2} shows the characterization of each type. In this table, types 1 - 3 are regular spectra. Types 4 - 6 have splicing problems. Type 8 and Type 9 have emission signals. A careful analysis of the features is detailed in Sect. \ref{sec:4-2}.      

\begin{table}[htp]
\centering
\bc
\renewcommand\arraystretch{1}	
	\caption{The basic information of each type.}
    \centering
	\label{tab1}
	\begin{tabular}{ccccccccc} 
		\hline           
	    Type & Type & \multirow{2}{*}{Number} & Proportion  & S/N & S/N &  Seeing & Seeing  & Lunar date  \\
      No. & name &   & (\%) &  average &  range  &   average&  range & mode \\
		\hline
1	&	 R-EL 	&	100136	&	15.70 	&	5.54 	&	[0,877.44]	&	3.53 	&	[0,9]	&	28	\\
2	&	 R-ML 	&	114954	&	18.02 	&	4.61 	&	[0,761.74]	&	3.58 	&	[0,9]	&	28	\\
3	&	 R-LL 	&	57671	&	9.04 	&	6.60 	&	[0,893.77]	&	3.51 	&	[0,9]	&	28	\\
4	&	 M-SS 	&	89638	&	14.05 	&	4.91 	&	[0,842.09]	&	3.49 	&	[0,9]	&	17	\\
5	&	 M-SU 	&	57109	&	8.95 	&	4.78 	&	[0,568.27]	&	3.47 	&	[0,9]	&	18	\\
6	&	 M-SA 	&	177290	&	27.79 	&	4.89 	&	[0,806.36]	&	3.54 	&	[0,9]	&	28	\\
7	&	 WF 	&	22413	&	3.51 	&	3.91 	&	[0,435.79]	&	3.48 	&	[0,9]	&	28	\\
8	&	 G-NNEs 	&	3244	&	0.51 	&	7.23 	&	[0,520.25]	&	3.53 	&	[1.9,9]	&	28	\\
9	&	 G-$\alpha$WE 	&	307	&	0.05 	&	5.42 	&	[1.98,90.93]	&	3.51 	&	2.1,5.9]	&	22	\\
10	&	 TER 	&	234	&	0.04 	&	4.29 	&	[1.48,65.65]	&	3.66 	&	[0,9]	&	18	\\
11	&	 SSE 	&	3329	&	0.52 	&	6.39 	&	[0,82.68]	&	3.74 	&	[2.4,4.8]	&	17	\\
12	&	 HBV 	&	10941	&	1.71 	&	42.58 	&	[0,998.37]	&	3.45 	&	[1.4,9]	&	29	\\
13	&	 OUF 	&	728	&	0.11 	&	9.81 	&	[0,173.06]	&	3.54 	&	[1.9,7.8]	&	25	\\

		\hline
	\end{tabular}
\ec
\end{table}

      \begin{table}[htp]
\centering
\bc
\renewcommand\arraystretch{1}	
	\caption{The characterization of each type.}
    \centering
	\label{tab2}
	\begin{tabular}{ll||ll} 
		\hline
	      Type name & Feature representation & Type name& Feature representation\\
		\hline
             R-EL 	&	 Regular-Early type-Low Quality 	&	 G-NNEs 	&	 Galaxy-Nebula-Narrow Emissions 	\\
             R-ML 	&	 Regular-Middle type-Low Quality 	&	 G-$\alpha$WE 	&	 Galaxy-Alpha-Wide Emission 	\\
             R-LL 	&	 Regular-Late type-Low Quality 	&	 TER 	&	 Telluric lines-Emissions-Residue 	\\
             M-SS 	&	 Mis-Splicing-Stagger 	&	 SSE 	&	 Specific Region-Specific Emissions 	\\
             M-SU 	&	 Mis-Splicing-Uncalibrated 	&	 HBV 	&	 Handicapped spectra with various Bad Values 	\\
             M-SA 	&	 Mis-Splicing-Arched 	&	 OUF 	&	 Outlier with Un-gregarious Features 	\\
             WF 	&	 With-Weak-Features 	&				
            \\					

		\hline
	\end{tabular}
\ec
\end{table}

\subsection{Characteristic analysis}\label{sec:4-2}
This subsection provides a detailed analysis of the spectral characteristics of each type.

\begin{figure}[ht]
\begin{minipage}[t]{0.48\linewidth}
    \flushright
    \includegraphics[scale=1.0]{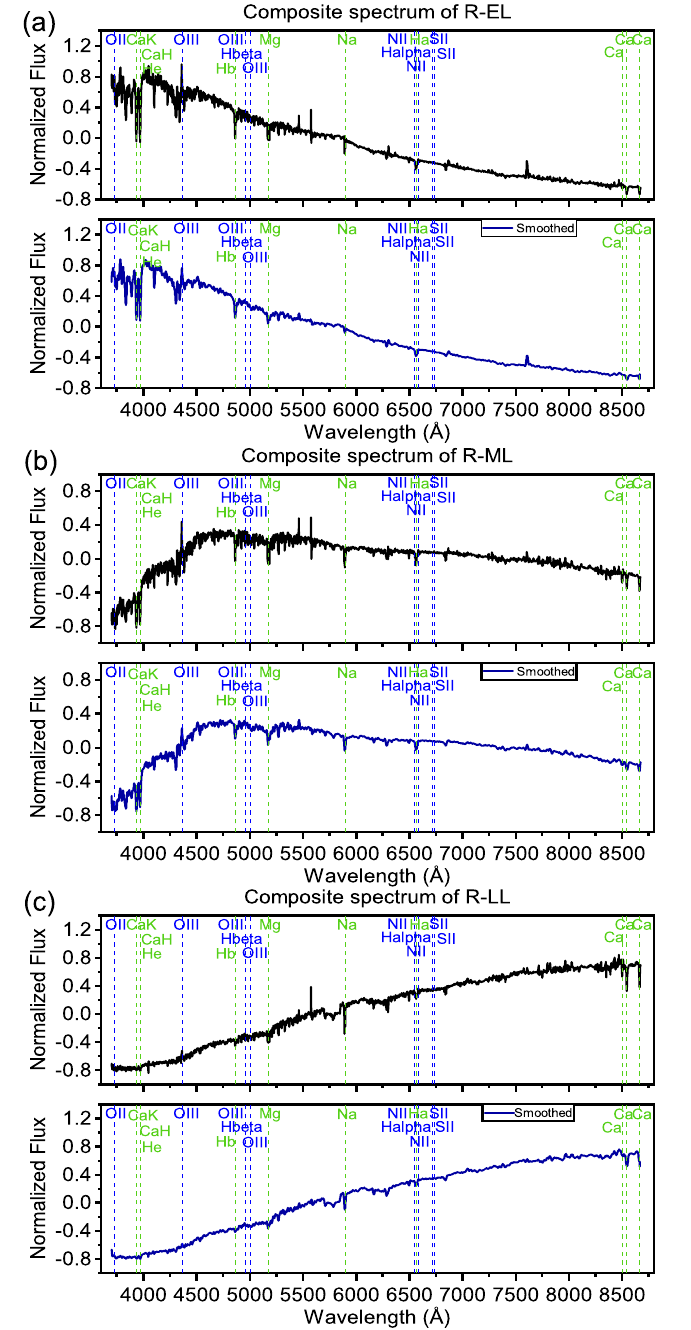}
    \caption{The composite spectra (top) of R-EL (a), R-ML (b) and R-LL (c) and smoothed spectra (bottom). The blue dotted lines indicate the emission lines and the green dotted lines indicate the absorption lines.} 
    \label{fig:example3} 
\end{minipage}%
    \hfill%
\begin{minipage}[t]{0.48\linewidth}
    \includegraphics[scale=1.0]{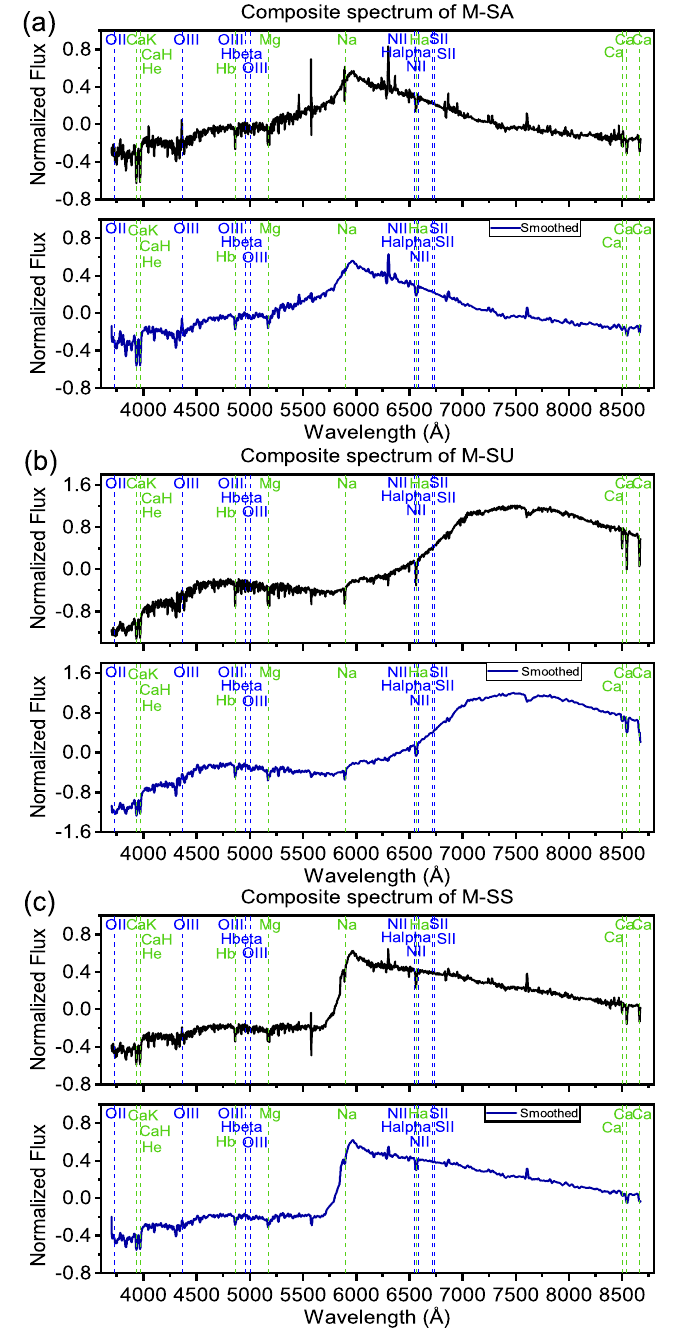}
    \caption{The composite spectra (top) of M-SS (a), M-SU (b) and M-SA (c) and smoothed spectra (bottom).}
    \label{fig:example4} 
\end{minipage} 
\end{figure}

$\textbf{Type1: R-EL}$ (Fig. \ref{fig:example3} (a)). The composite spectrum of this type exhibits a decreasing continuum trend from the blue end to the red end, showing obvious CaII H$\&$K, He and weak H$\beta$, Mg, Na, H$\alpha$ absorption line characteristics. These spectral features suggest that the targets may belong to early-class stars (class B, class A, and class F). Therefore, this type is named R-EL (Regular-Early type-Low Quality).

$\textbf{Type2: R-ML}$ (Fig. \ref{fig:example3} (b)). The composite spectrum of this type exhibits an increasing and then slowly decreasing trend in its continuum, and displays insignificant absorption lines, such as CaII H\& K, He, H$\beta$, H$\alpha$, CaII triplet lines, while the Mg and Na metal lines are slightly strong. It is speculated that the targets in this type may belong to intermediate classes (class F, class G and class K) between early-class and late-class stars. Therefore, this type is named R-ML (Regular-Middle type-Low Quality).

$\textbf{Type3: R-LL}$ (Fig. \ref{fig:example3} (c)). The composite spectrum of this type is characterized by an upward trend in its continuum, and almost no absorption line characteristics, while Na and CaII triplet lines are obvious. The spectra of this type are speculated to belong to late-class stars (class K and class M). Therefore, this type is named R-LL (Regular-Late type-Low Quality).

The common characteristics of above three types are that these spectra all show normal stellar spectral characteristics, enabling the determination of the corresponding stellar spectral subclass. At the same time, the stellar subclasses to which the spectra belong in the above three types have no obvious boundaries, resulting in crossover when judging the subclasses of early-class, medium-class and late-class stars. 

$\textbf{Type4: M-SS}$ (Fig. \ref{fig:example4} (a)). The composite spectrum of this type is characterized by a clear flux dislocation in the wavelength range of 5700$\AA$ -- 6000$\AA$, and some weak absorption line characteristics, such as CaII H$\&$K, He, H$\beta$, Mg, Na, H$\alpha$, and CaII triplet lines. Since the composite spectrum of this type presents a staggered shape, it is named M-SS (Mis-Splicing-Stagger).

$\textbf{Type5: M-SU}$ (Fig. \ref{fig:example4}(b)). The composite spectrum of this type is characterized by pumps at the blue band and red band, showing weak CaII H$\&$K, He, H$\beta$, Mg, Na and H$\alpha$ absorption lines and stronger CaII triplet lines. Since the features of this type are uncalibrated, it is named M-SU (Mis-Splicing-Uncalibrated). 

$\textbf{Type6: M-SA}$ (Fig. \ref{fig:example4} (c)). The composite spectrum of this type is characterized by an upward trend at the blue band and a downward trend at the red band of the spectrum. The overall shape of the spectrum is arched, with the peak occurring at the wavelength around 6000$\AA$. Its line characteristics are similar to M-SS but extremely noisy. Because the composite spectrum of this type is arched, it is named M-SS (Mis-Splicing-Arched).

The common characteristics of above three types are that all of them suffer from poor splicing and have very weak line characteristics. 

\begin{figure}[h]
\centering
\includegraphics[scale=1.0]{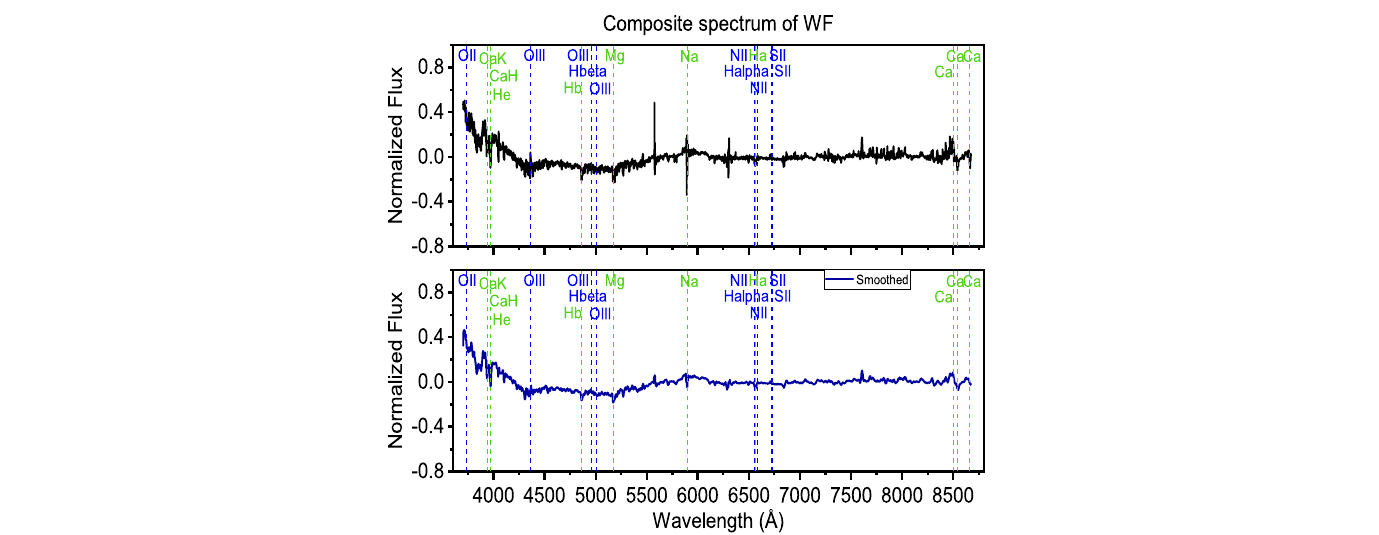}
\caption{The composite spectrum (top) of WF and smoothed spectrum (bottom). } 
\label{fig:example5}
\end{figure}

$\textbf{Type7: WF}$ (Fig. \ref{fig:example5}). The composite spectrum of this type is characterized by a weak downward trend at the blue band and a flat trend at the red band of the spectrum. Only the weak CaII H\& K absorption line characteristics can be observed in the spectrum, but the wavelength of the above two lines is less than 4000$\AA$ with low confidence. Wavelengths beyond 4000$\AA$ are covered by noise where line characteristics are barely visible. Due to the weak features exhibited in the spectrum, this type is named WF (With-Weak-Features).

\begin{figure}[h]
\centering
\includegraphics[scale=1.0]{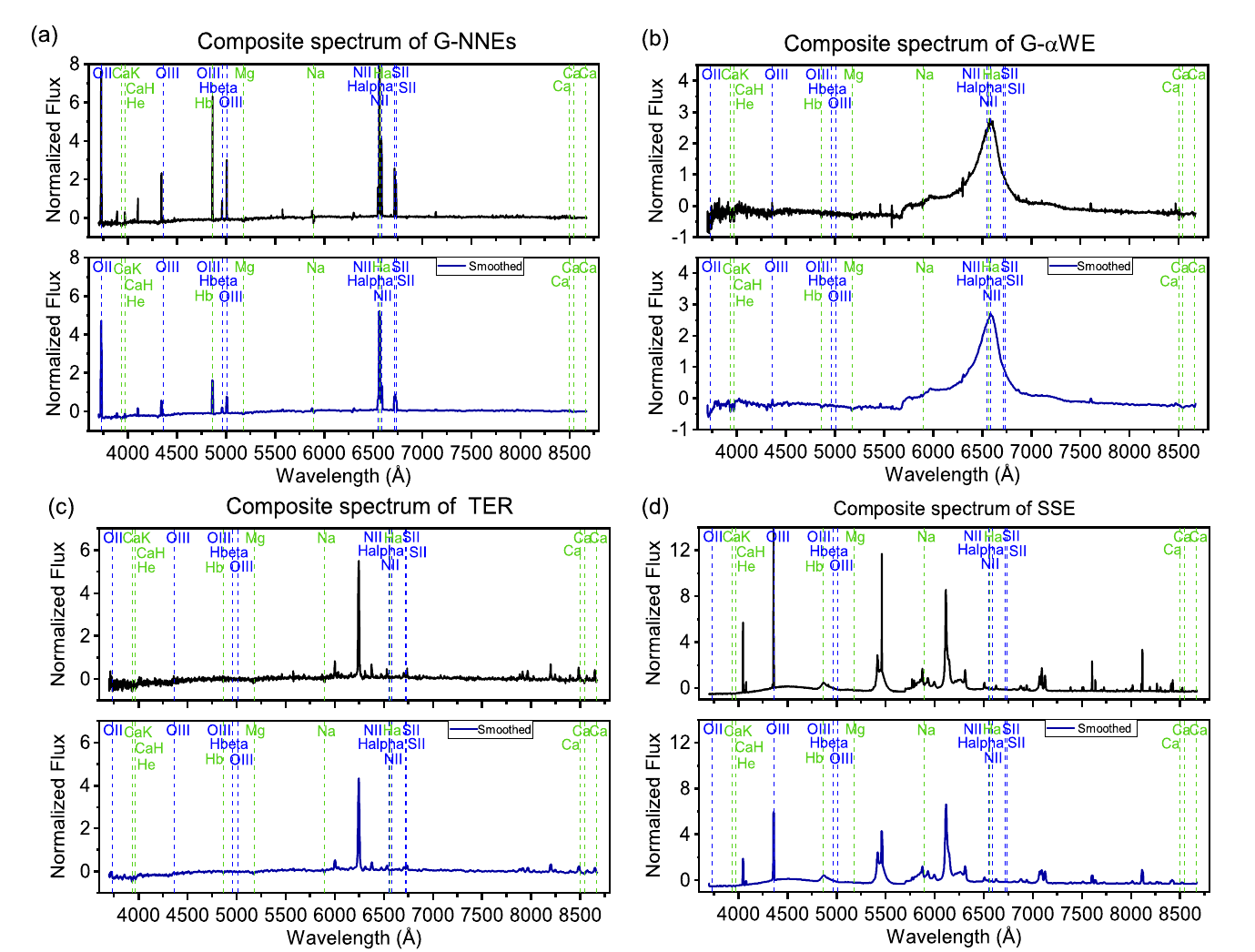}
\caption{The composite spectra (top) of G-NNEs (a), G-$\alpha$WE (b), TER (c) and SSE (d) and smoothed spectra (bottom).} 
\label{fig:example6}
\end{figure}

$\textbf{Type8: G-NNEs}$ (Fig. \ref{fig:example6} (a)). The composite spectrum of this type is characterized by nebular emission lines at rest wavelength, such as OII, OIII, H$\beta$, NII, H$\alpha$, and SII, which show strong 0-redshift. Due to the spectrum exhibiting galactic nebular features and narrow emission lines, it is named G-NNEs(Galaxy-Nebula-Narrow Emissions).

\begin{figure}[ht]
\begin{minipage}[t]{0.48\linewidth}
    \flushright
    \includegraphics[scale=1.0]{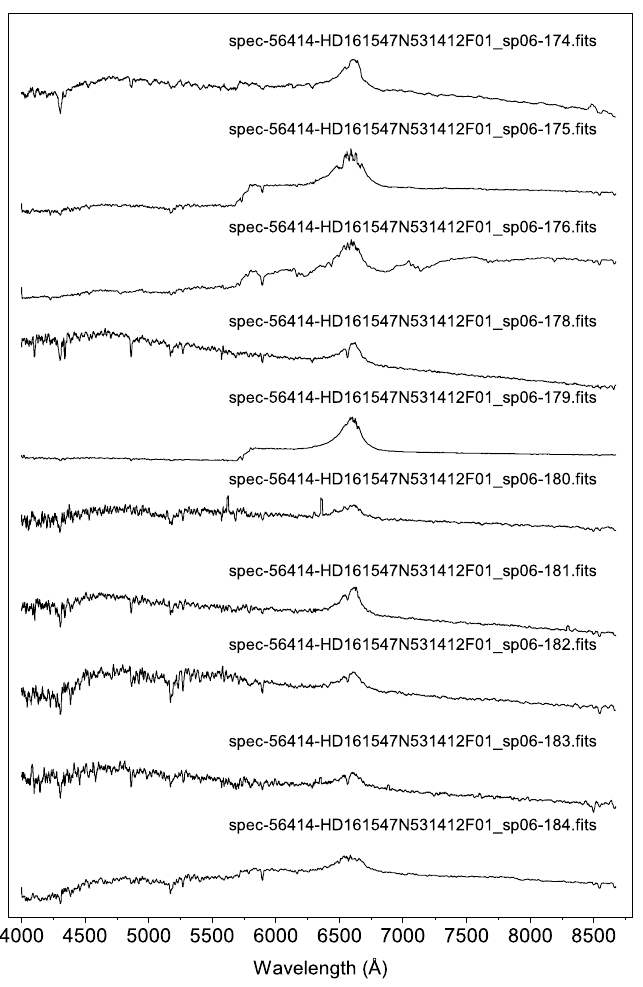}
    \caption{The adjacent fibers spectra of G-$\alpha$WE, from the plate HD161547N531412F01 spectrograph sp06 on May 1, 2013. The abnormal emission features are visible in the 6564$\AA$ range.} 
    \label{fig:example7} 
\end{minipage}%
    \hfill%
\begin{minipage}[t]{0.48\linewidth}
    \includegraphics[scale=1.0]{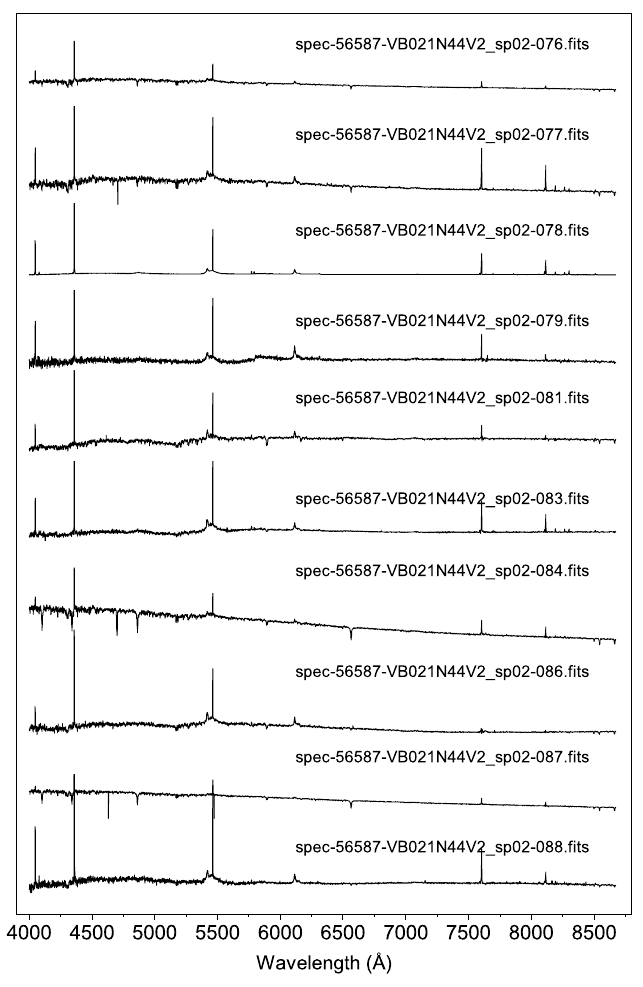}
    \caption{The adjacent fibers spectra of SSE, from the plate VB021N44V2 spectrograph sp02 on October 21, 2013. The strong line characteristic is the mercury line in 5460$\AA$.}
    \label{fig:example9} 
\end{minipage} 
\end{figure}

$\textbf{Type9: G-$\pmb{\alpha}$WE}$ (Fig. \ref{fig:example6} (b)). The composite spectrum of this type is characterized by a strong and broad emission line suspected H$\alpha$ in the rest wavelength, but no other emission line characteristics were found. Moreover, we have noticed that this type of feature happens on adjacent fibers(Fig. \ref{fig:example7}), implying a light contamination origin. Corresponding to G-NNEs, this type is named G-$\pmb{\alpha}$WE (Galaxy-Alpha-Wide Emission).



$\textbf{Type10: TER}$ (Fig. \ref{fig:example6} (c)). The composite spectrum of this type is characterized by an extremely strong emission line at a wavelength of about 6250$\AA$, which is a telluric line characteristic. Faint SII emission lines can be observed in the rest of the band, which may be residual components of the sky subtraction. Therefore, this type is named TER (Telluric lines-Emissions-Residue). The number of this type is very small, accounting for about 0.04\% of the total `Unknown'.

$\textbf{Type11: SSE}$ (Fig. \ref{fig:example6} (d)). The composite spectrum of this type is characterized by a set of strong emission features, and the position of this set of strong emission features in the spectrum is fixed. We found that observation dates are concentrated on October 21, 2013 and December 26, 2016, and observation targets are only in four specific regions. Therefore, this type is named SSE (Specific Region-Specific Emissions). To further explore the mechanism behind the spectral features of the SSE type,  we check the other spectra of adjacent fibers on the same date (examples are shown in Fig. \ref{fig:example9}). On all of the spectra, the contamination of the mercury lamp emission lines (from city lights) at 5460$\AA$ is clearly seen.

\begin{figure}[h]
\centering
\includegraphics[scale=1.0]{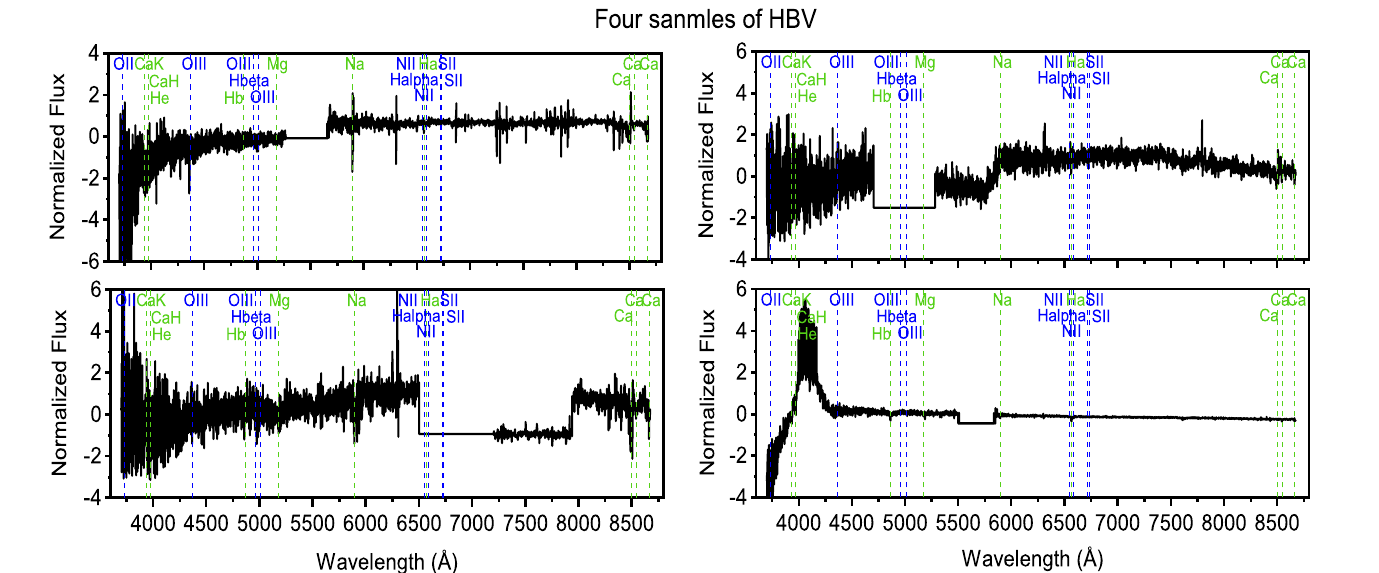}
\caption{Four spectral samples randomly selected in HBV. } 
\label{fig:example10}
\end{figure}

\begin{figure}[h]
\centering
\includegraphics[scale=1.0]{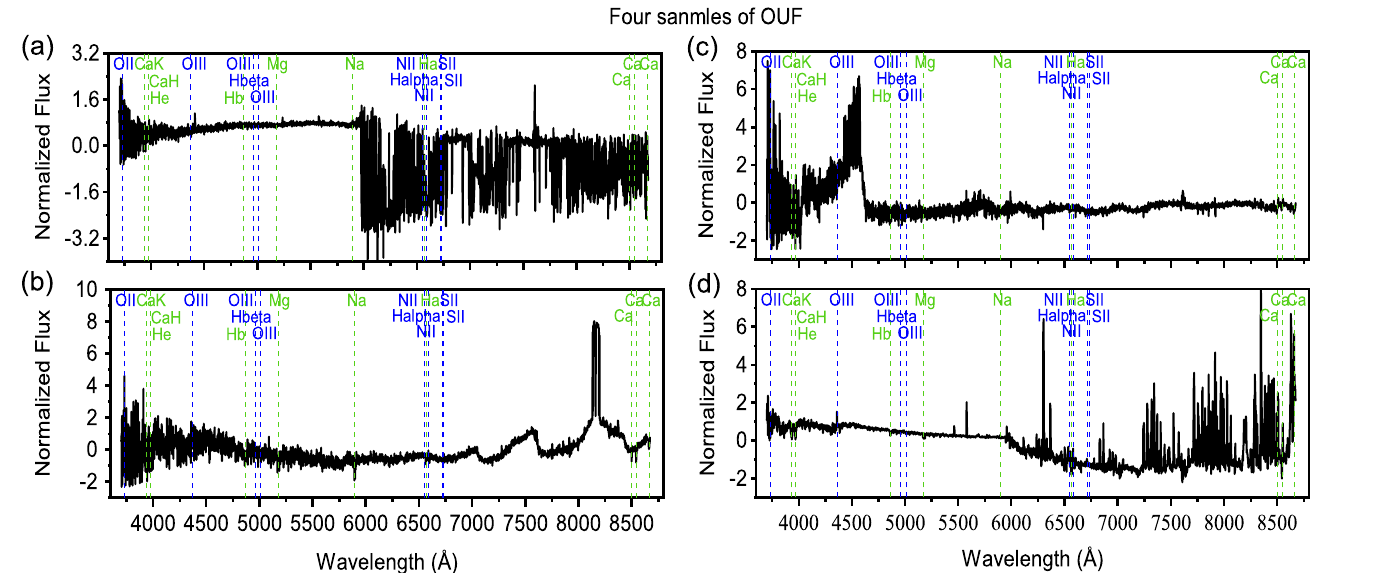}
\caption{Four spectral samples randomly selected in OUF.} 
\label{fig:example11}
\end{figure}

$\textbf{Type12: HBV}$ (Fig. \ref{fig:example10}). The common characteristic of the spectra in HBV is that these spectra present one or more segments of continuous bad pixel values and the locations are diverse \citep{kang2021novel}. Therefore, this type is named HBV (Handicapped spectra with various Bad Values).

$\textbf{Type13: OUF}$ (Fig. \ref{fig:example11}). This type collects out-of-type spectra (d\textgreater 3$\sigma$) from each of the previous types. The spectra in this type have no common features, and they are too different from the features of the other types to get a match. Therefore, this type is named OUF (Outlier with Un-gregarious Features). The main reason why these spectra are classified as `Unknown' is that each spectrum presents one or more very strong pseudo features that are poorly correlated with the observed target. As shown in Figure \ref{fig:example11}, for the four spectral examples in this type, their spectra features are: strong burrs at the red wavelengths (a), strong abnormal emission line around 8200$\AA$ (b), strong jumps around 4500$\AA$ (c), and strong residual sky emission lines at the red wavelength (d) respectively. We believe that these abnormal features greatly affect the identification of the target spectrum. On the other hand, we think peculiar spectra are more likely to appear in this type because they are un-gregarious to any other types.

\section{Origin analysis of `Unknown' spectra}\label{sec:4}
Many factors may lead to poor matching of a spectrum to the known templates, such as low quality data, incomplete templates, peculiar spectral features, contamination from various sources, problems of data processing, etc. In order to reveal the causes behind the `Unknown' spectra and provide valuable information for the follow-up observations and data processing, this section explores the origins of these 13 types clustered by NAPC-Spec algorithm from different perspectives. 

\subsection{Spectral signal-to-noise ratio}

\begin{figure}[htbp]
\centering
\includegraphics[scale=0.9]{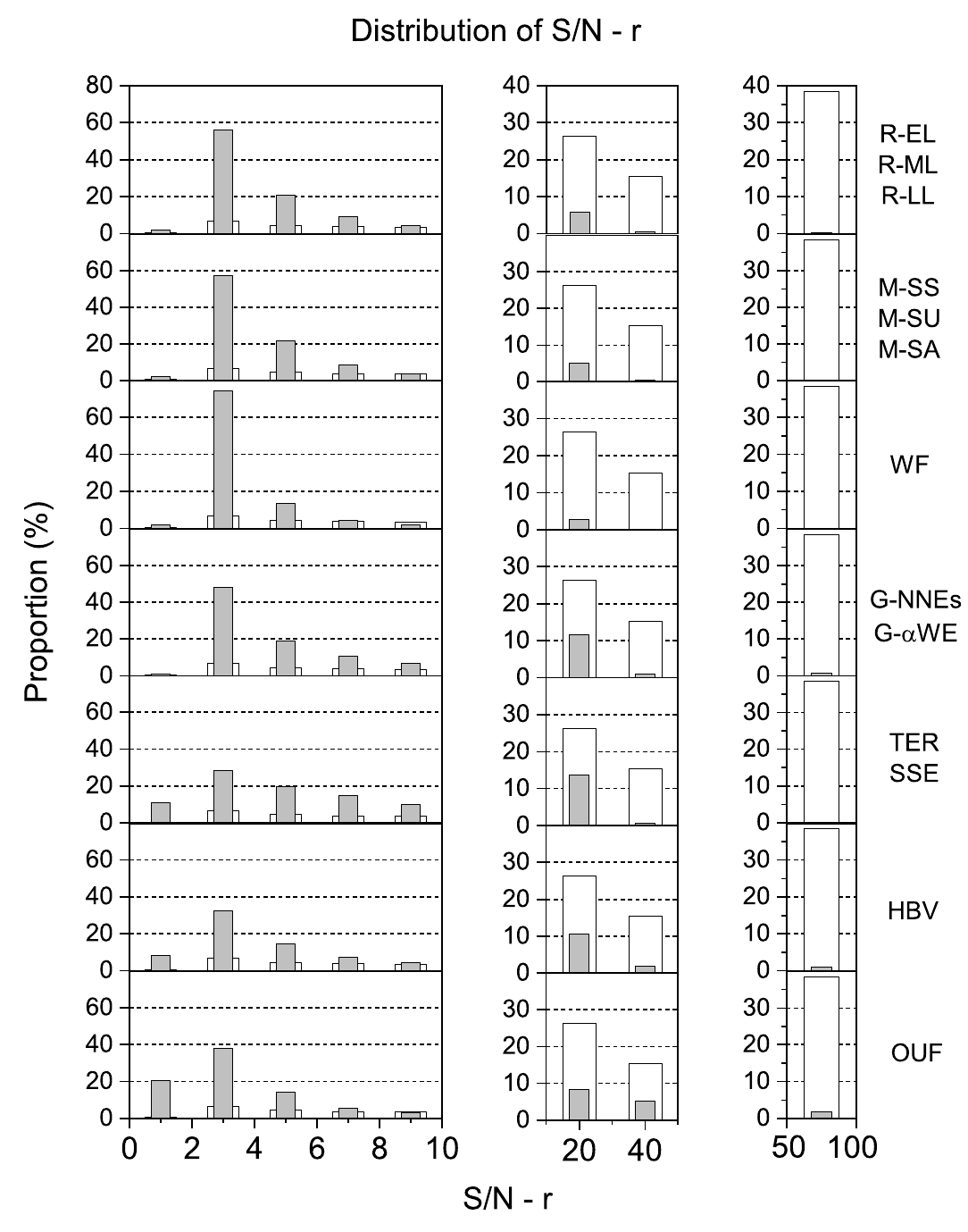}
\caption{Distribution of spectral S/N. The white histograms show the percentage distribution of all DR5 spectra in each interval, and the gray histograms show the percentage distribution of different types (type names are shown on the right side). Since the S/N of the `Unknown' spectra is mainly concentrated in (0, 10], the step size is set to 2, the step size is set to 20 in (10, 50], and the step size is set to 50 in (50, 100).} 
\label{fig:example13}
\end{figure}

We first show the S/N distributions of various types of ‘Unknown’ spectra in Figure \ref{fig:example13}. The S/N value is represented by the average S/N in \textit{r} band wavelength. The average S/N values and S/N ranges of each type are also listed in Table \ref{tab1}.

The average S/N of all DR5 spectra is 45.54, while the average S/N of `Unknown' spectra is only 5.25. As can be seen from the white histograms of Figure \ref{fig:example13}, 80\% of all DR5 spectra have S/N greater than 10, which is much higher than the `Unknown' spectra (gray histograms). In particular, the S/N-r of [R-EL R-ML R-LL], [M-SS M-SU M-SA] and WF are extremely low and mainly concentrated in the [2, 4] interval, and [G-NNEs G-$\alpha$WE], [TER SSE], HBV and OUF have a roughly uniform S/N distribution.

\subsection{Footprint distribution of the targets}
\begin{figure}[h]
\centering
\includegraphics[width=140mm]{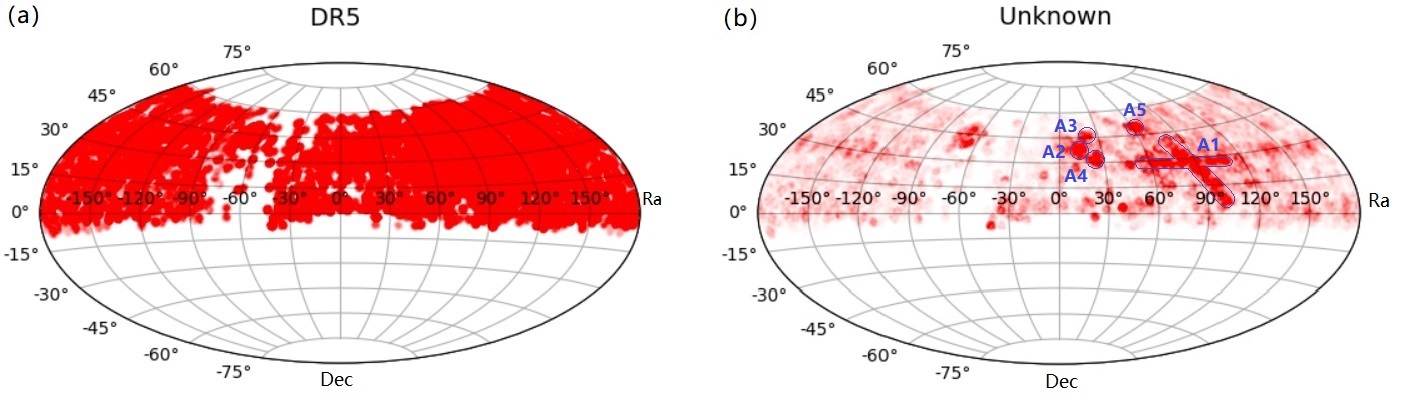}
\caption{The footprint comparison between `Unknown'(Panel b) and all DR5 targets(Panel a).} 
\label{fig:example12}
\end{figure}

Figure \ref{fig:example12} shows the footprint distribution comparison between ‘Unknown’ and all DR5 targets. There are five local high-density areas in Figure \ref{fig:example12} (b), which are one `+'-like area (noted as A1) at RA range: $60^{\circ}$ - $120^{\circ}$, and DEC range: $0^{\circ}$ - $45^{\circ}$, and four circle areas with radius $\approx 2.5^{\circ}$ (corresponding to the field of view of one LAMOST plate). The centre coordinates (RA, DEC) of these four circle areas are  ($14^{\circ}$, $37^{\circ}$), ($20^{\circ}$, $46^{\circ}$), ($22^{\circ}$, $31^{\circ}$) and ($62^{\circ}$, $50^{\circ}$) (noted as A2, A3, A4, and A5) respectively.

As can be seen, the `Unknown' spectra mainly concentrate in M31 and Galactic Anti-centre region (GAC) \citep{xiang2017lamost}.
In fact, we find high percentages of the targets from M31 member stars in A2 (78.8\%), A3 (55.3\%) and A4 (79.7\%) respectively \citep{chen2015lamost}. The environment of A1 is complex. About 4\% members of the well-known Pleiades cluster, a high percentage of variable stars and other small clusters are located in this area \citep{jayasinghe2018asas,samus2017general,mowlavi2021large,heinze2018first,tian2020catalog,chen2020zwicky,sampedro2017multimembership,hernitschek2017geometry,delchambre2019gaia}.  Not only that, many of the LAMOST median-depth plates and faint  plates are located at these footprints. The faint magnitude of the target sources in these plates may be one of the main reasons for the concentration of the 'Unknown' spectra in these footprints.

As shown by the footprint, the distribution of `Unknown' spectra is also related to the observational season. Compared with all DR5 observed objects, the proportion of `Unknown' objects is 8.2\% in spring, 29.4\% in summer, 6.8\% in autumn and 7.8\% in winter. Obviously, the percentage of `Unknown' objects from the summer is higher than other seasons.

In all DR5 objects, the `Unknown' objects account for 7.0\% in the low Galactic latitudes (l \textless 45°) and 7.4\% in the high silver latitudes (l \textgreater 45°). It indicates that the `Unknown' ratio is similar between extra-galactic and galactic targets.

\subsection{Lunar phase}

\begin{figure}[h]
\centering
\includegraphics[scale=1]{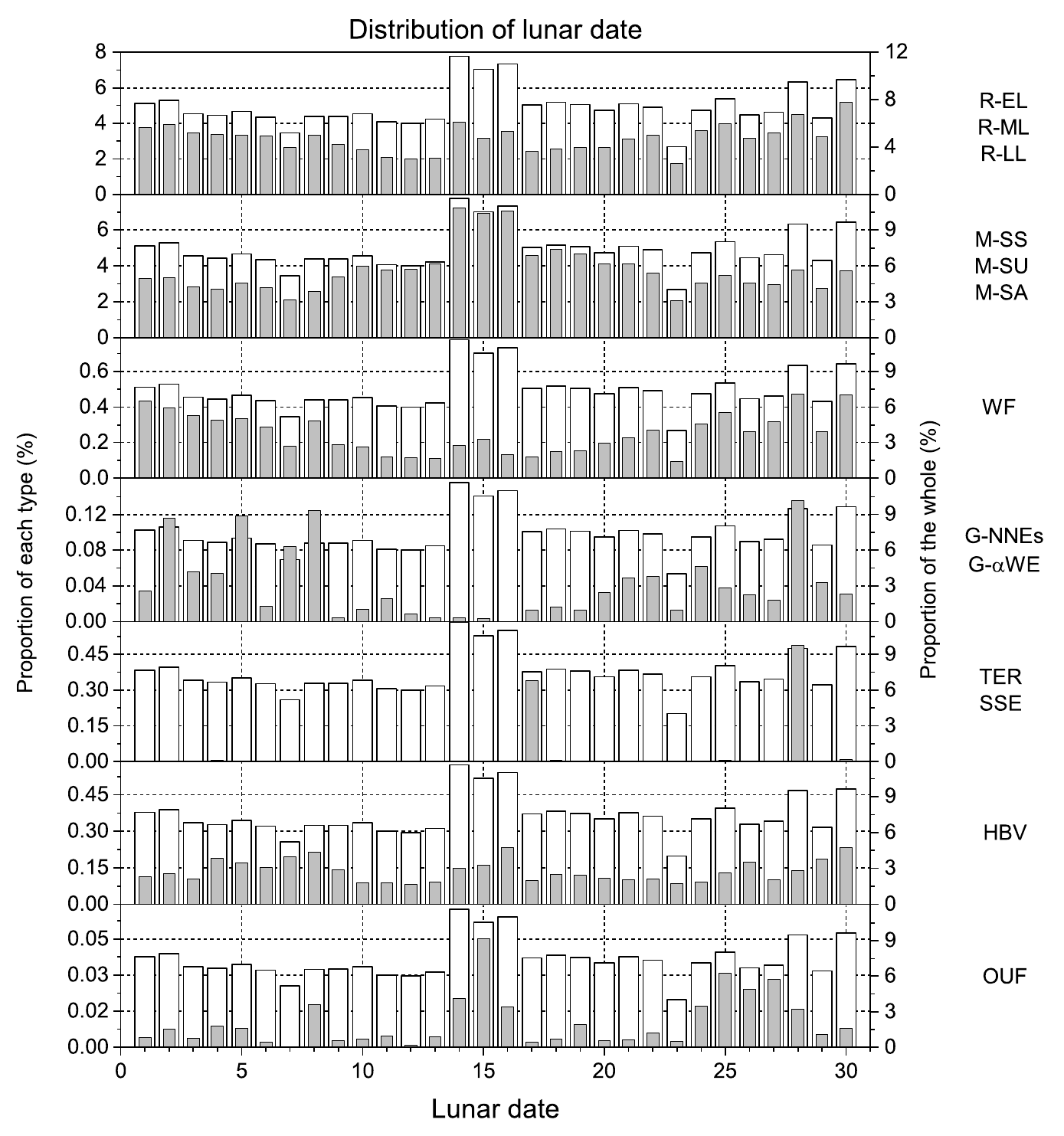}
\caption{Lunar date distribution for `Unknown' spectral observations. The white histograms (sub-coordinates) show the proportional distribution of the number of observations of all `Unknown' spectra for different lunar dates versus the number of all DR5 spectra. The gray histograms (primary coordinates) show the proportional distribution of the number of observations of different types for different lunar dates versus the number of all DR5 spectra (types name is displayed on the right). } 
\label{fig:example14}
\end{figure}

Sky subtraction \citep{bai2017sky} is a necessary step of the LAMOST spectral processing. However, the fibers assigned for background are not enough to build a super sky, especially on bright nights \citep{luo2015first}, which may result in moonlight residues on the target spectra. To further investigate the effect of lunar phase on data quality, the distribution of observation lunar dates for the `Unknown' spectra is presented in Figure \ref{fig:example14}.

It can be seen from the white histograms of Figure \ref{fig:example14} and Table \ref{tab1} that the proportion of the `Unknown' spectra is higher during bright moon nights. It is consistent with the result in \citep{yang2020tad,WOS:000466450300049}. The number of `Unknown' spectra on the lunar calendar 7 and 23 of every month is the smallest, and these days are assigned as the test days for some specific celestial targets. The proportion of [R-EL R-ML R-LL], [M-SS M-SU M-SA] and OUF on the lunar calendar 14, 15, and 16 in every month is relatively large, and their distributions are consistent with the whole `Unknown' spectra. 


\subsection{Seeing}

\begin{figure}[htbp]
\centering
\includegraphics[scale=1]{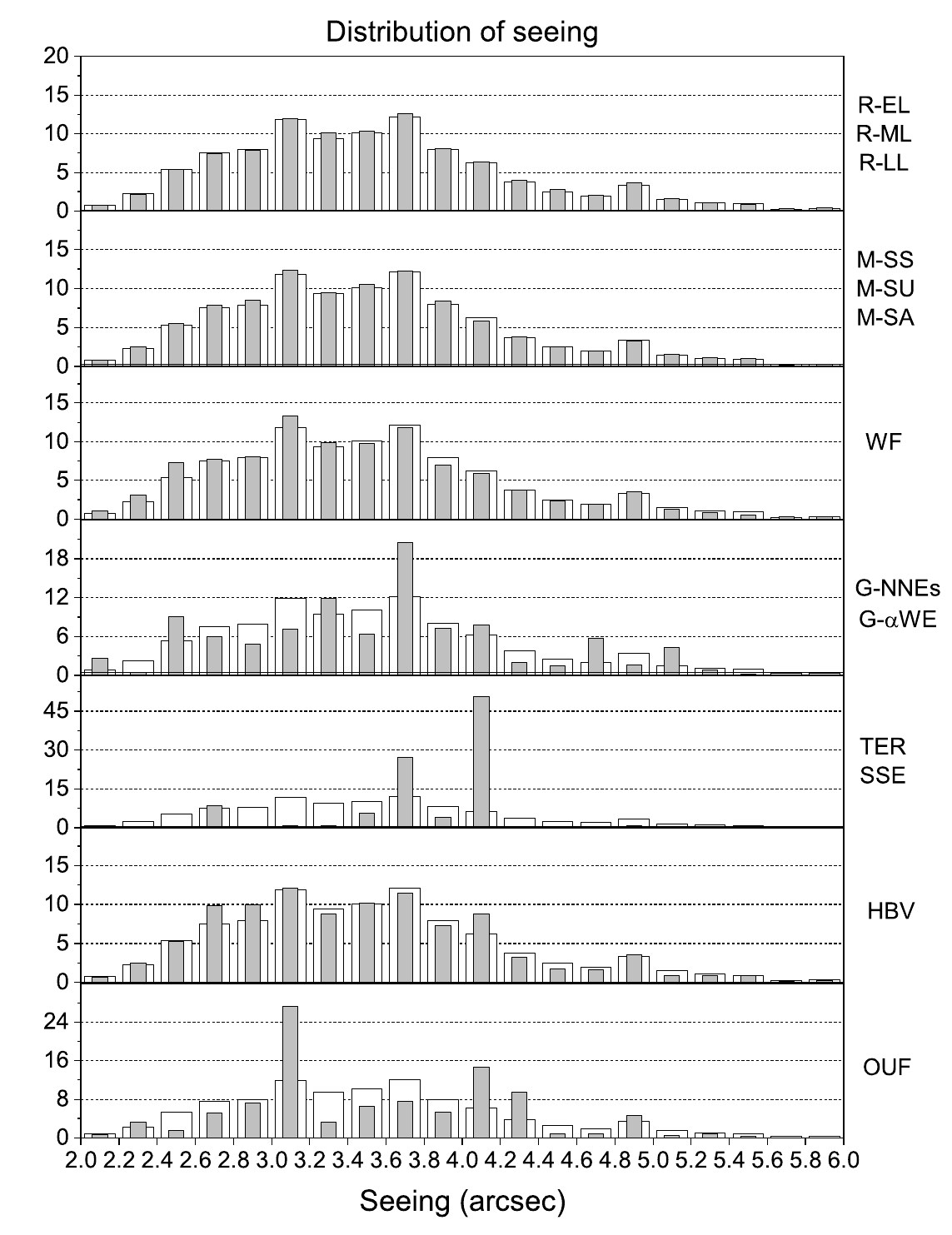}
\caption{Seeing distribution of the `Unknown' spectra. The white histograms show the proportional distribution of the number of all `Unknown' spectra in each interval of seeing, and the gray histograms show the proportion of the number of different types in each interval of seeing (type names are listed on the right). The distribution in intervals (0,2.0) and (6.0, $\infty$) is not shown in this figure since the number of spectra contained in the above intervals is few.} 
\label{fig:example17}
\end{figure}

The seeing at the time of the observation of the LAMOST spectral survey has been calculated by manually measuring the full width at half maximum of guide star image. We show the seeing distribution of the `Unknown’ spectra in Figure \ref{fig:example17}.

The white histograms show that most of the seeing values are distributed in [2.4,4.4], with a trend roughly obeysing the Gaussian distribution and peaks at [3.0,3.8]. However, the average value of LAMOST data is around 1.9 arcsec. Thus, the `Unknown' spectra have relatively high seeing values. Observing the gray histograms of different types in Figure \ref{fig:example17}, the distribution characteristics of [R-EL R-ML R-LL], [M-SS M-SU M-SA], WF and HBV are basically consistent with the white histograms. 

\begin{figure}[htbp]
\centering
\includegraphics[scale=1.0]{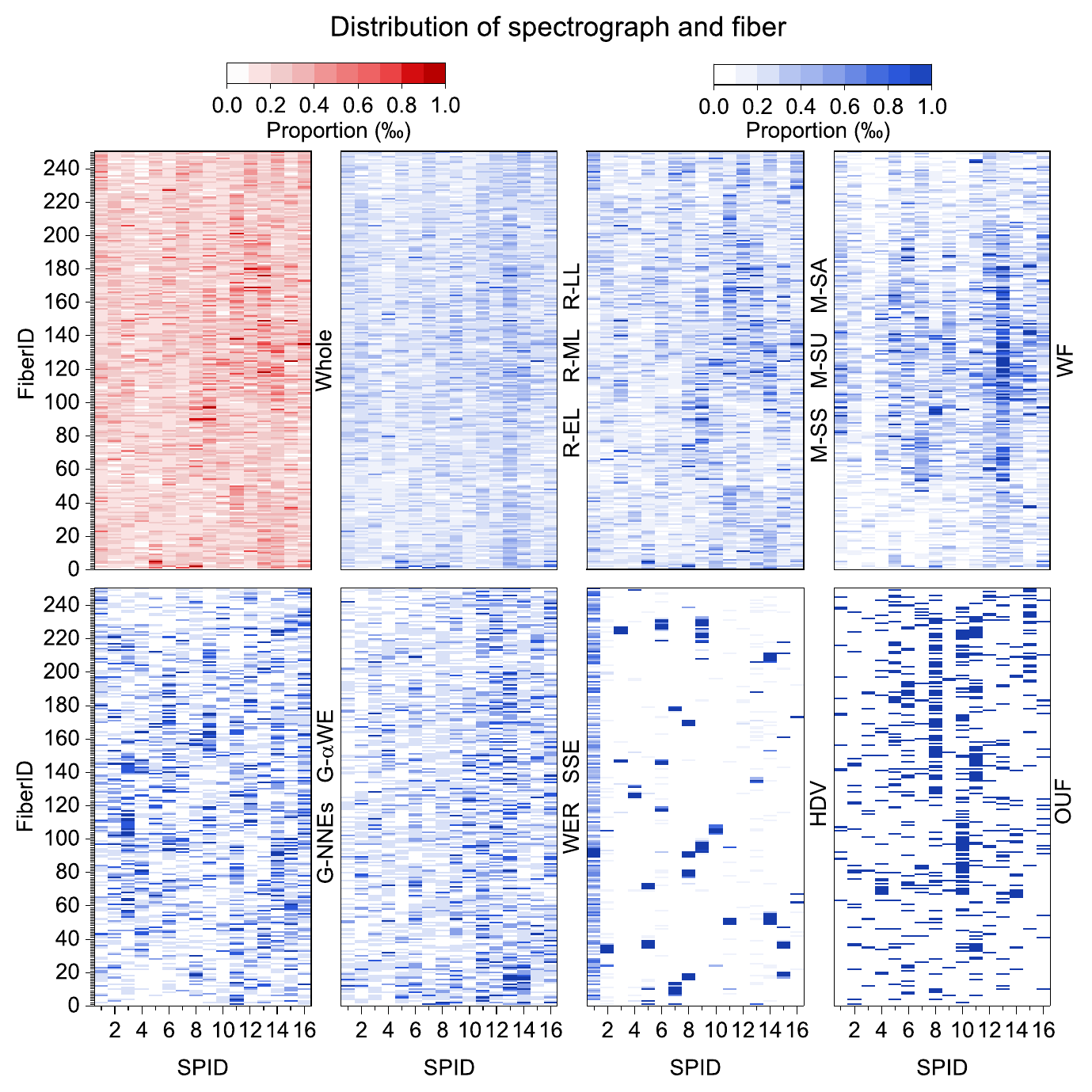}
\caption{Spectrographs and fibers distribution of `Unknown' spectra. The figure in red gives the distribution of all `Unknown' spectra in each spectrograph and in each fiber (denoted as Whole), and the figures in blue present the distributions of different types (type names are listed on the right) in each spectrograph and each fiber.} 
\label{fig:example18}
\end{figure}

\subsection{Working status of spectrograph and fiber}
The quality of the spectra may be affected by the working status of the spectrographs and the fibers. Figure \ref{fig:example18} shows the distribution of the `Unknown' spectra on the instruments labeled as 1-16 in spectrograph and 1-250 in fiber.

As for the whole `Unknown' sample, they are concentrated on the 110th to 200th fibers of the 9th, 11th, 12th, 13th and 14th spectrographs. The distributions of [R-EL R-ML R-LL], [M-SS M-SU M-SA], WF and [TER SSE] are similar to the whole sample, while the distributions of [G-NNEs G-$\alpha$WE] are relatively scattered. For HBV, it is mainly distributed in the fibers of 1st spectrograph.


\section{Discussion}\label{sec:4}
In this work, we developed SA-Frame, a methodological framework specifically designed for the systematic analysis of `Unknown' spectra. The NAPC-Spec clustering analysis makes it possible not only to discover targets of rare features but also to separately explore their underlying mechanisms. The NAPC-Spec algorithm has good resistance to noise and degeneracy in high-dimensional data. Nevertheless, a small number of spectra are still quite different from the composite spectra after clustering. The generation of the un-gregarious type effectively enhanced the cleanliness of all other type. 

All the `Unknown' spectra of LAMOST DR5 are clustered into 13 types according to their density distribution, an analysis and characterization of each type are given according to continuum shapes and line characteristics. On this basis, origin analysis is carried out for each type from the perspectives of observational conditions and the working status of the instruments. In addition, many other factors from data processing may affect the data quality, including dark and bias subtraction, flat field correction, spectral extraction, sky subtraction, wavelength calibration, merging sub-exposures, combining wavelength bands \citep{luo2015first,luo_zhang_chen_song_wu_bai_wang_du_zhang_2013,luo2004design}. These factors cannot be easily traced backward, exploring these factors of data processing on the 'Unknown' spectra would be valuable for future work.


%

\begin{acknowledgements}
The authors thank the reviewer (Professor Shiyin Shen) for the many suggestions that have helped to improve the manuscript.\\
The work is supported by the National Natural Science Foundation of China (Grant Nos. U1931209, 62272336), Projects of Science and Technology Cooperation and Exchange of Shanxi Province (Grant Nos. 202204041101037, 202204041101033), and the central government guides local science and technology development funds (Grant No. 20201070). The Fundamental Research Program of Shanxi Province (Grant Nos. 20210302123223, 202103021224275), and the PhD Start-up Foundation of Taiyuan University of Science and Technology(20221007).\\
Guoshoujing Telescope (the Large Sky Area Multi-Object Fiber Spectroscopic Telescope, LAMOST) is a National Major Scientific Project built by the Chinese Academy of Sciences. Funding for the project has been provided by the National Development and Reform Commission. LAMOST is operated and managed by the National Astronomical Observatories, Chinese Academy of Sciences.
\end{acknowledgements}




\bibliographystyle{raa}
\bibliography{ref}

\label{lastpage}

\end{document}